\documentclass[aps,prb,preprint,showpacs,floatfix,superscriptaddress]{revtex4-1}
\usepackage{graphicx,epsfig,dcolumn,amssymb,amsmath}
\usepackage{booktabs}
\usepackage{float}
\usepackage{color}

\begin{document}

\title{Stable Ultra-thin CdTe Crystal: A Robust Direct Gap Semiconductor}

\author{F. Iyikanat}
\email{fadiliyikanat@iyte.edu.tr}
\affiliation{Department of Physics, Izmir Institute of Technology, 35430, Izmir, Turkey}

\author{B. Akbali}
\affiliation{Department of Physics, Izmir Institute of Technology, 35430, Izmir, Turkey}

\author{J. Kang}
\affiliation{Materials Sciences Division, Lawrence Berkeley National Laboratory, Berkeley, California 94720, United States}

\author{R. T. Senger}
\affiliation{Department of Physics, Izmir Institute of Technology, 35430, Izmir, Turkey}
\affiliation{ICTP-ECAR Eurasian Center for Advanced Research, Izmir Institute of Technology, 35430, Izmir, Turkey}

\author{Y. Selamet}
\affiliation{Department of Physics, Izmir Institute of Technology, 35430, Izmir, Turkey}

\author{H. Sahin}
\email{hasansahin@iyte.edu.tr}
\affiliation{ICTP-ECAR Eurasian Center for Advanced Research, Izmir Institute of Technology, 35430, Izmir, Turkey}
\affiliation{Department of Photonics, Izmir Institute of Technology, 35430, Izmir, Turkey}

\date{\today}
\date{\today}

\begin{abstract}

Employing density functional theory based calculations, we investigate structural, vibrational and 
strain-dependent electronic properties of an ultra-thin CdTe crystal structure that can be derived from its 
bulk counterpart. It is found that this ultra-thin crystal has an 8-atom primitive unit cell with considerable 
surface reconstructions. Dynamic stability of the structure is predicted based on its calculated vibrational 
spectrum. Electronic band structure calculations reveal that both electrons and holes in single layer CdTe 
possess anisotropic  in-plane 
masses and mobilities. Moreover, we show that the ultra-thin CdTe has some interesting electromechanical 
features, such as strain-dependent anisotropic variation of the band gap value, and its rapid increase under 
perpendicular compression. The direct band gap semiconducting nature of the ultra-thin CdTe crystal remains 
unchanged under all types of applied strain. With a robust and moderate direct band gap, single-layer CdTe is 
a promising material for nanoscale strain dependent device applications.

\end{abstract}

\maketitle

\section{Introduction}

Discovery of graphene\cite{Novoselov} has attracted great interest towards the family of two-dimensional (2D) 
crystal structures. In  addition to graphene, other 2D crystals such as germanene,\cite{Cahangirov, Bianco} 
silicene,\cite{Vogt, Sahin} stanene,\cite{Zhu, Rachel} transition-metal dichalcogenides (TMDs),\cite{Wang, Esfahani, 
Island, Chuang, Kang} and post-transition-metal chalcogenides (TMCs)\cite{Late, Cai,  Yagmurcukardes} have been 
predicted and successfully synthesized. One of the most prominent members of 2D crystals is MoS$_{2}$. 
MoS$_{2}$ exhibits a transition from an indirect band gap of 1.29 eV to a direct band gap of 1.90 eV when its 
layer thickness is reduced from bulk to a single-layer.\cite{Mak, Splendiani} Single-layer MoS$_{2}$ based 
field-effect transistors (FETs) can have room-temperature on/off ratios of the order of 10$^{8}$ and these 
fabricated transistors can exhibit a carrier mobility larger than 200 cm$^{2}$$/$(V s).\cite{Radisavljevic, 
Radisavljevic1} Moreover, MoS$_{2}$ has excellent mechanical properties like high flexibility\cite{Pu, Lee} 
and high strength.\cite{Bertolazzi} Due to these outstanding properties, 2D materials will play an important 
role in the applications of future
optoelectronics and flexible electronics.

Recent studies have shown that not only layered materials but also ultra-thin forms of non-layered materials 
that consist of a few atomic layer thickness can form 2D crystals.\cite{Tan, Vogt, Bacaksiz} For 
instance, CdSe, CdS and CdTe nanoplatelets with thicknesses ranging from 4 to 11 monolayers were 
synthesized.\cite{Ithurria} The thickness dependence of the absorption and emission spectra of these 
nanoplatelets were demonstrated. Park \textit{et al.} achieved successful synthesis of 1.4-nm-thick ZnSe 
nanosheets with wurtzite structure.\cite{Park} Using a colloidal template method large-scale fabrication of 
free-standing ultra-thin and lamellar-structured CdSe with wurtzite crystal structure was achieved.\cite{Son} 
Furthermore, using a lamellar hybrid intermediate, large-area, free-standing, single-layers of ZnSe were 
fabricated.\cite{Sun} Single-layers of ZnSe-pa (pa stands for n-propylamine) were exfoliated from a lamellar 
hybrid (Zn$_{2}$Se$_{2}$)(pa) intermediate. Then, by heat treatment pa-molecules were cleared off and the 
colloidal suspension of clean ZnSe single-layers was obtained. Fabricated single-layer ZnSe has 
four-atomic-layer thickness. They showed that, produced single-layer ZnSe was highly stable over several days. 
The photocurrent densities of these monolayers are much higher than that of their bulk counterparts.

Cadmium telluride (CdTe) is one of the most popular II-VI semiconductors because of its potential applications 
in optoelectronic devices such as  photodetectors, solar cells and room temperature X- and gamma-ray 
detectors.\cite{Rogalski, Gupta, Tu, Szeles} CdTe has a direct optical band gap of $\sim$ 1.5 eV with a high 
absorption coefficient.\cite{Ferekides, Mahabaduge} Solar cell efficiency of CdTe-based thin-films has 
recently reached 22.1\%.\cite{Green} CdTe crystallizes in the zinc-blende structure at room temperature. The 
CdTe thin films can be grown by various deposition techniques such as chemical vapor deposition,\cite{Kim} 
pulsed laser deposition,\cite{Diamant}  electrochemical deposition\cite{Bonilla} and spray 
pyrolysis.\cite{Ison} Generally, the intrinsic properties of ultra-thin materials exhibit drastic changes 
compared to their bulk counterparts. Thus, when a material is thinned from bulk to ultra-thin form, it can 
exhibit enhanced properties and new functionalities.

In this study, motivated by the recent synthesis of ultra-thin II-VI binary compounds, we investigate structural, 
electronic and vibrational properties of single-layer CdTe using first principle calculations based on density 
functional theory (DFT). Although there are a few prior computational studies on single-layer 
CdTe,\cite{Wang1, Zheng} free-standing monolayer CdTe has not been predicted yet. We found that single-layer 
CdTe containing eight atoms in the primitive unit cell is structurally stable with anisotropic electronic 
properties. It has a direct band-gap at the $\Gamma$ point and direct gap transition at the $\Gamma$ point is 
not affected by strain along any direction. The strain-dependent anisotropic variation of the band gap value 
and its rapid increase under out-of-plane compression pressure are found. The paper is organized as follows: 
details of the computational methodology are given in Sec. II. Structural and electronic properties of 
single-layer CdTe are presented in Sec. III. In Sec. IV, the dynamical stability of single-layer CdTe is 
studied. Effect of strain on electronic properties is discussed in Sec. V. Finally, we outline our results in 
Sec. VI.

\begin{figure}
\includegraphics[width=10 cm]{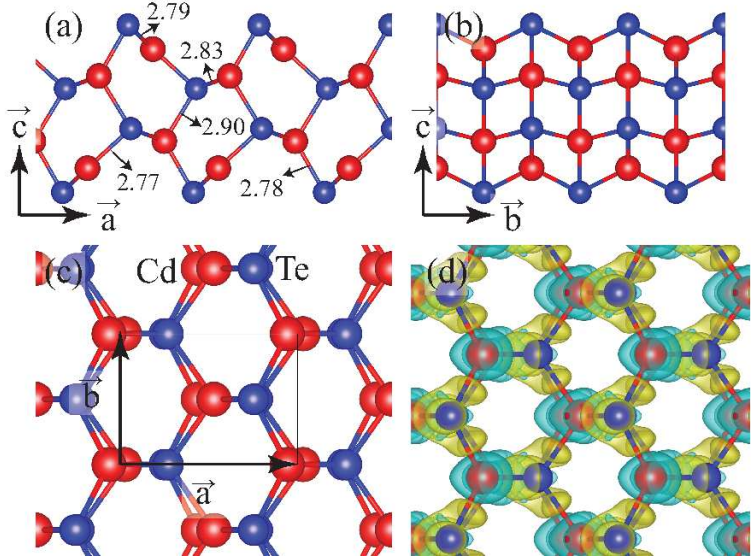}
\caption{\label{structure1}
(Color online)
Side views (a) along $\vec{a}$ lattice vector, (b) along $\vec{b}$ lattice vector and (c) top view of single-layer CdTe.
Black lines represent the rectangular unit cell.
(d) The charge densities of the isolated Cd and Te atoms are subtracted from the charge density of 
single-layer CdTe. The yellow and blue densities stand for the negative and positive charges, respectively. 
Red and blue atoms are for Cd and Te, respectively.}
\end{figure}

\section{Computational Methodology}

All calculations were performed within the density functional theory (DFT) using projector-augmented-wave 
potentials (PAW)  and a plane-wave basis set as implemented in the Vienna \textit{ab initio} simulation 
package (VASP).\cite{Kresse, Kresse1} Perdew-Burke-Ernzerhof (PBE) version of generalized gradient 
approximation (GGA)\cite{Perdew} was used for the description of the exchange-correlation functional. Analysis 
of the charge transfers in the structures was made by the Bader technique.\cite{Henkelman} The ionization 
energy is determined as the energy difference between the valance band maximum energy and the vacuum level at the (110) 
side of the bulk and single-layer CdTe.

The conjugate gradient algorithm was used to optimize the structure. The cutoff energy for the plane-waves 
was chosen to be 500 eV. The convergence criterion for energy was taken to be 10${^{-5}}$ eV between two 
consecutive steps. The convergence for the Hellmann-Feynman force in each unit cell was taken to be 10$^{-4}$ 
eV$/$\AA{}. The pressure in the unit cell was kept below 1 kBar. In order to eliminate  interlayer interaction 
within the periodic images, a vacuum spacing of approximately 12 \AA{} between adjacent layers was chosen. For 
the structural optimization, a 9$\times$12$\times$1 $\Gamma$-centered \textit{k}-point mesh was used. The 
cohesive energy per atom was calculated using the formula \begin{equation}\label{eq}
E_{c}=[ n_{Cd}E_{Cd}+n_{Te}E_{Te}-E_{SL}] / n \end{equation} where E$_{Cd}$ and E$_{Te}$ are isolated single 
atom energies for Cd and Te, respectively. While $n$ stands for the number of all atoms, $n_{Cd}$ and $n_{Te}$ 
show the numbers of Cd and Te atoms in the unit cell, respectively. E$_{SL}$ denotes the total energy of the 
single-layer CdTe. Phonon dispersions and eigenvectors are calculated by making use of the small displacement 
methodology implemented in the PHON code.\cite{Alfe}

\begin{figure}
\includegraphics[width=10 cm]{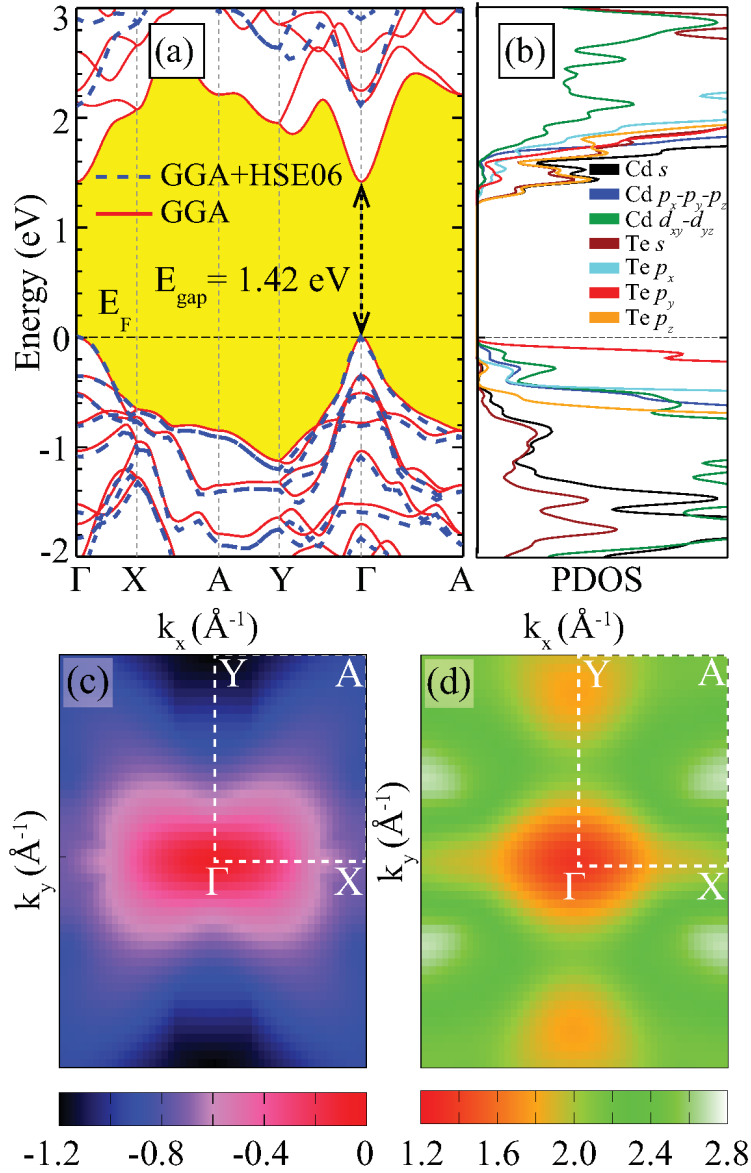}
\caption{\label{structure2}
(Color online)
(a) Electronic band structure and (b) atom- and orbital-decomposed electronic density of states of single-layer CdTe. 
Fermi level is set to zero. (c) and (d) 2D surface plots of the valence band and the conduction band edges in the 
reciprocal space, respectively. The energy values (eV) are color coded below the plots.}
\end{figure}

\begin{table}[htbp]
	\caption{\label{table1}The calculated ground state properties for bulk and single-layer (SL) CdTe: The 
lattice constants, \textit{a} and \textit{b}; atomic distance between Cd and Te atoms, d$_{Cd-Te}$; charge transfer 
from Cd to Te atom, $\Delta\rho$; the cohesive energy per atom, E$_{c}$; energy band gap, E$_{gap}$; and ionization 
energy, I. E.}
\begin{tabular}{lcccccccc}
	\hline\hline
	& $a$     & $b$   &  d$_{Cd-Te}$ & $\Delta\rho$  &E$_{c}$ & E$_{gap}$ & I. E. \\
	& (\AA{}) &(\AA{})   &   (\AA{})       &($\textit{e}$) & (eV)   & (eV) & (eV)\\
	\hline
	Bulk      CdTe       & 6.52 &  -   &    2.82       &0.5 & 2.20 & 0.72 & 5.22 \\
	SL CdTe      & 6.18 &  4.53   &  2.77-2.90       & 0.5 & 1.79 & 1.42 & 5.15\\
	\hline\hline
\end{tabular}
\end{table}

\section{Structural and Electronic Properties of Single-Layer C\lowercase{d}T\lowercase{e}}


It is well-known that the bonding character of zinc-blende CdTe is partly covalent and partly 
ionic.\cite{Groiss, Guo} Except for the (110) facets, zinc-blende structure of CdTe has polar surfaces, 
which are chemically highly active. Even if single-layer structures having these polar surfaces could be 
obtained, their chemical activity would hinder their stability. However, since the (110) surfaces are 
non-polar, cleavage along these planes could be more feasible. 

The proposed structure of CdTe single-layers in our study have the same crystal structure as the fabricated 
highly stable single-layer of zinc-blende ZnSe (\textit{zb}-ZnSe).\cite{Sun} Side views along $\vec{a}$ and 
$\vec{b}$ directions and top view of single-layer CdTe are shown in Figs. \ref{structure1} (a)-(c), 
respectively. Lattice parameters of single-layer CdTe are found to be \textit{a} $=$ 6.18 and \textit{b} 
$=$ 4.53 \AA{}. Calculated lattice parameters are smaller than those for bulk CdTe which is 6.52 \AA{}. Fig. 
\ref{structure1} (a) shows that the Cd-Te bond lengths vary from 2.77 to 2.90 \AA{}, bond lengths 
between surface atoms being smaller than those of the inner atoms in the layer. As seen in Fig. 
\ref{structure1} (b), Te atoms are at the surfaces of the layer, and each surface Te atom binds to three Cd 
atoms. The inner Te atoms are surrounded by 4 Cd atoms with tetrahedral type bonds. During the atomic 
relaxation of the truncated layer, Cd atoms that are at the surface recede toward the inner Te atoms; 
remaining Te atoms move outward. Such reconstructions stabilize the layer surfaces.

Bader charge analysis reveals that each Cd atom donates 0.5\textit{e} to each Te atom. To illustrate the 
charge transfer mechanism three dimensional charge density differences are shown in Fig. \ref{structure1} 
(d). The charge density differences were calculated by subtracting charge of isolated Cd and Te atoms from 
charge of single-layer CdTe. The charge transfer between Cd and Te atoms resembles polar-covalent bonding. Due 
to a difference in the electronegativities of Cd and Te atoms (1.69 and 2.10 for Cd and Te atoms, 
respectively), the Cd-Te bonding has also some ionic character. Finally, the cohesive energy per atom of 
single-layer CdTe is 1.79 eV which is less than  the bulk value of 2.20 eV per atom.

\begin{figure}
\includegraphics[width=11 cm]{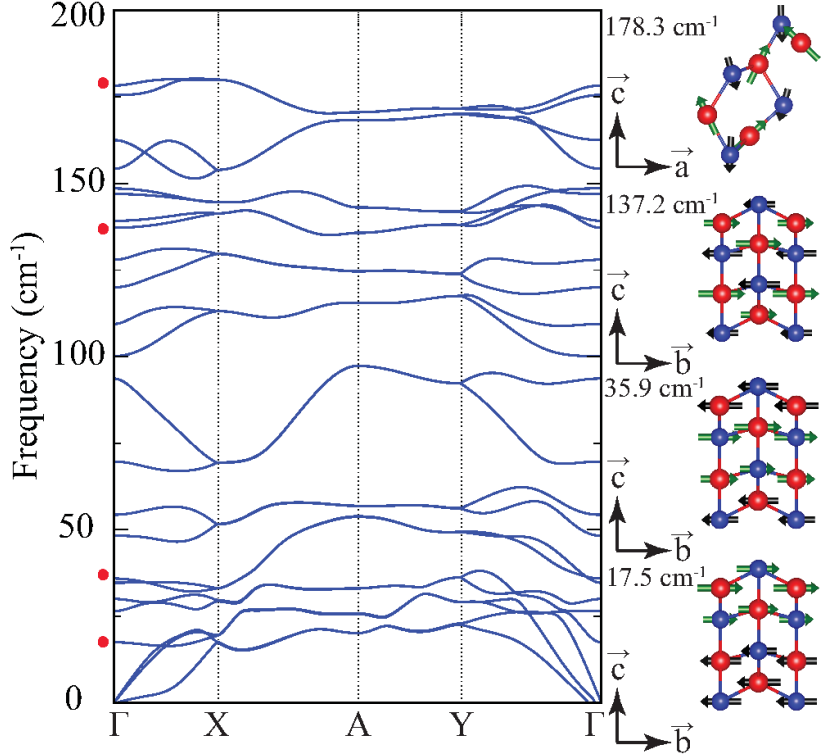}
\caption{\label{structure3}
(Color online) The phonon spectrum of monolayer CdTe is shown on the left. The branches of the possible 
Raman-active modes are indicated by the red dots and the corresponding normal modes are shown on the right.}
\end{figure}


To investigate the full band dispersions and the characteristics of band edges in the Brillouin Zone (BZ), 
whole BZ energy-band structure is calculated and given in Fig. \ref{structure2} (a). As shown in the figure, 
valence band maximum (VBM) and conduction band minimum (CBM) of CdTe reside at the same symmetry point of the 
$\Gamma$. The calculated GGA electronic-band structure demonstrate that single-layer CdTe is a direct-gap 
semiconductor with a band gap of 1.42 eV. In order to give more accurate gap energy of the single-layer 
CdTe the calculated band structures within HSE06 correction are also shown Fig. \ref{structure2} (a). A 
calculated HSE06 gap of single-layer CdTe is 2.13 eV. Since the trend and qualitative behavior of all the 
bands calculated using GGA and HSE06 are similar, only the GGA based results are given in the rest of the 
paper.

\begin{table*}[htbp]
\caption{\label{table2} Effective masses of electrons (m$_{e}$) and holes (m$_{h}$) of bulk and single-layer (SL) CdTe.}
\begin{tabular}{lcccccccc}
\hline\hline
      & m$_{e}$ ($\Gamma\rightarrow$L) & m$_{e}$ ($\Gamma\rightarrow$X) & m$_{e}$ ($\Gamma\rightarrow$Y) & m$_{e}$ ($\Gamma\rightarrow$A) & m$_{h}$ ($\Gamma\rightarrow$L) & m$_{h}$ ($\Gamma\rightarrow$X) & m$_{h}$ ($\Gamma\rightarrow$Y) & m$_{h}$ ($\Gamma\rightarrow$A) \\
\hline
Bulk CdTe & 0.10 & 0.09 & - & - & 0.84 & 0.72 & - & - \\
SL CdTe  & - & 0.39 & 0.17 & 0.21 & - & 0.74 & 0.14 & 0.20  \\
\hline\hline
\end{tabular}
\end{table*}

In order to properly understand the electronic properties of CdTe, partial density of states (PDOS) is also 
plotted in Fig. \ref{structure2} (b). The states in the vicinity of VBM are mostly composed of 
\textit{p$_{y}$} orbitals of Te. These \textit{p$_{y}$} orbitals of Te atoms are parallel to the \textit{b} 
lattice vector of the unit cell. On the other hand, CBM is mostly made up of the \textit{s} orbitals of Cd and 
the \textit{s} and \textit{p$_{z}$} orbitals of Te. Note that, the \textit{p$_{z}$} orbital contribution of Te 
atom mainly comes from surface Te atoms. Two-dimensional contour plots of the valence band (VB) and the 
conduction band (CB) of the single-layer CdTe are shown in Figs. \ref{structure2} (c) and (d). The directional 
anisotropy at the band edges is clearly seen in the surface plots. Ionization energy (I. E.) 
of single-layer and bulk CdTe surfaces are also calculated and are shown in Table \ref{table1}.

Due to reduced crystal symmetry in a single layer form of a material, its electronic 
characteristics are quite different from their bulk forms. Moreover, in-plane anisotropy in the ultra-thin 
materials can lead to significant modifications in the electronic properties of the material. Therefore, the 
investigation of direction-dependent electronic properties of ultra-thin materials is of importance. The 
effective masses of electron (m$_{e}$) and hole (m$_{h}$) of single-layer CdTe are calculated near the 
$\Gamma$ point. Our calculations show that the m$_{e}$ and m$_{h}$ effective masses are highly anisotropic 
around the $\Gamma$ point. As given in Table \ref{table2} m$_{e}$ values are 0.39, 0.17 and 0.21 for 
$\Gamma$$\rightarrow$X, $\Gamma$$\rightarrow$Y and $\Gamma$$\rightarrow$A, respectively. m$_{h}$ values are 
0.74, 0.14 and 0.20 for $\Gamma$$\rightarrow$X, $\Gamma$$\rightarrow$Y and $\Gamma$$\rightarrow$A, 
respectively. As seen in Fig. \ref{structure2} (b), the VBM is mainly composed of p$_{y}$ electrons of Te 
atoms, thus this causes a high in-plane anisotropy in m$_{h}$ values. The anisotropy in the electron and hole 
masses are evident even from the crystal structure where \textit{x}- and \textit{y}-directions are highly 
anisotropic (see Fig. \ref{structure1}). For comparison, the calculated values of m$_{e}$ and m$_{h}$ of 
bulk CdTe are also given in Table \ref{table2}.

\begin{figure*}
\includegraphics[width=17 cm]{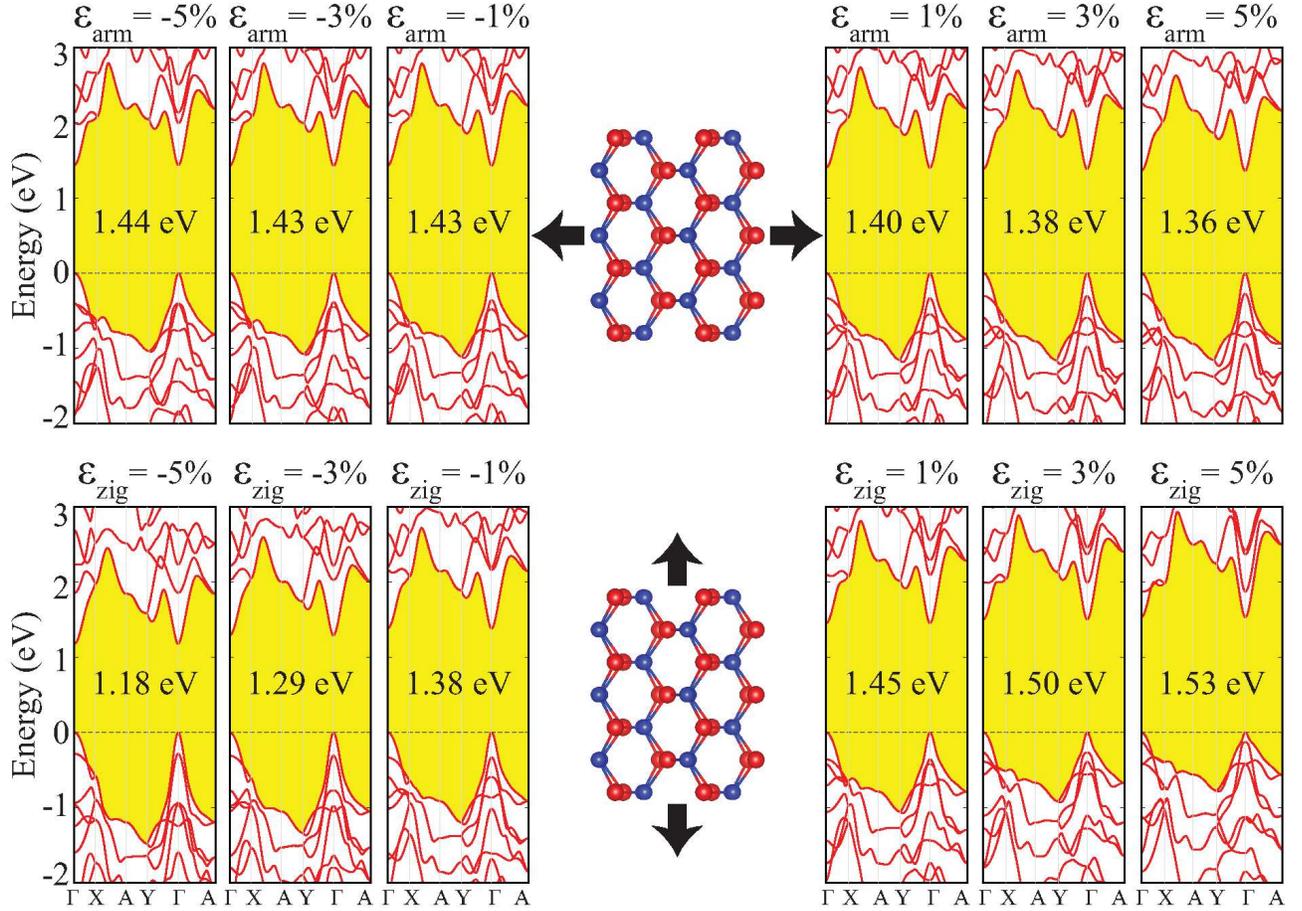}
\caption{\label{structure4}
(Color online) Evolution of the band dispersion of single-layer CdTe as a function of in-plane strain along 
armchair ($\varepsilon_{arm}$) and zigzag ($\varepsilon_{zig}$) directions. Fermi level is set to zero.}
\end{figure*}

\section{Dynamical Stability of Single-Layer C\lowercase{d}T\lowercase{e}}

Dynamical stability of the single-layer CdTe is investigated by examining the phonon spectra of the crystal. 
The small displacement method as implemented in the PHON software package is used to calculate the phonon 
spectra.\cite{Alfe} 4$\times$4$\times$1 supercell is used for the phonon-band structure calculations. In Fig. 
\ref{structure3}, we present the calculated phonon-band structure of single-layer CdTe obtained by the method 
described above. It is found that all the phonon modes have real eigenfrequencies, which indicate that CdTe 
single-layers are stable. The small imaginary frequencies (less than 1 cm$^{-1}$) near the $\Gamma$ point are 
numerical artifacts caused by the inaccuracy of the FFT grid and they get cured as larger and larger 
supercells are considered.

The structural characteristics of bulk \textit{zb}-CdTe were well studied in earlier Raman studies. The unit 
cell of bulk \textit{zb}-CdTe consists of one Cd and one Te atoms, therefore the phonon dispersion of bulk 
CdTe yields three acoustic and three optical modes. Main Raman active phonon modes are transverse optical (TO) 
and longitudinal optical (LO) modes and they occur approximately at 141 and 168 cm$^{-1}$.\cite{Amirtharaj} In 
addition to these prominent modes, A$_{1}$ and E symmetry modes were reported at 92, 103, 120 and 147 
cm$^{-1}$ which give information about the presence of Te on the surface of bulk CdTe.\cite{Amirtharaj, 
Zitter}

On the other hand, the unit cell of single-layer CdTe contains four Cd atoms and four Te atoms. Therefore, the 
phonon dispersion of single-layer CdTe possesses three acoustic and twenty-one optical modes as shown in Fig. 
\ref{structure3}. As pointed out in the previous section there is a relaxation of the top atomic layers in the 
single-layer CdTe and bond length of the surface atoms is shorter than bond length of the inner atoms. 
Distortions of surface atoms lead to several flat phonon bands in Fig. \ref{structure3}. These distortions 
lift the degeneracies at the $\Gamma$ point and lead to hybridization of the acoustic and optical phonon 
branches. Optical character and frequency of possible Raman active modes are shown in the right panel of Fig 
\ref{structure3}. The modes at 17.5, 35.9 and 137.2 cm$^{-1}$ have in-plane character (E$_{g}$ like) and the 
motion of the atoms are parallel to the $\vec{b}$. For the mode 137.2 cm$^{-1}$, Cd and Te atoms move in 
opposite directions. However, atomic layers exhibit contour-phase motion for the modes 17.5 and 35.9 
cm$^{-1}$. The mode with the highest frequency of 178.3 cm$^{-1}$ has mixed in-plane and out-of-plane 
character (A$_{g}$ like) with Cd and Te atoms having counter-phase motion.

Due to the heavier atomic masses and more ionic electronic character, phonon modes of single-layer CdTe lie at 
much lower energies than phonon modes of other 2D materials such as graphene, hBN and TMDs. Moreover, 
it was reported that phonon modes of structurally similar material of single-layer ZnSe lie at more higher 
energies than that of single-layer CdTe.\cite{Bacaksiz} Thus, it is clear that single-layer CdTe is a quite 
soft material.

\begin{figure}
\includegraphics[width=11 cm]{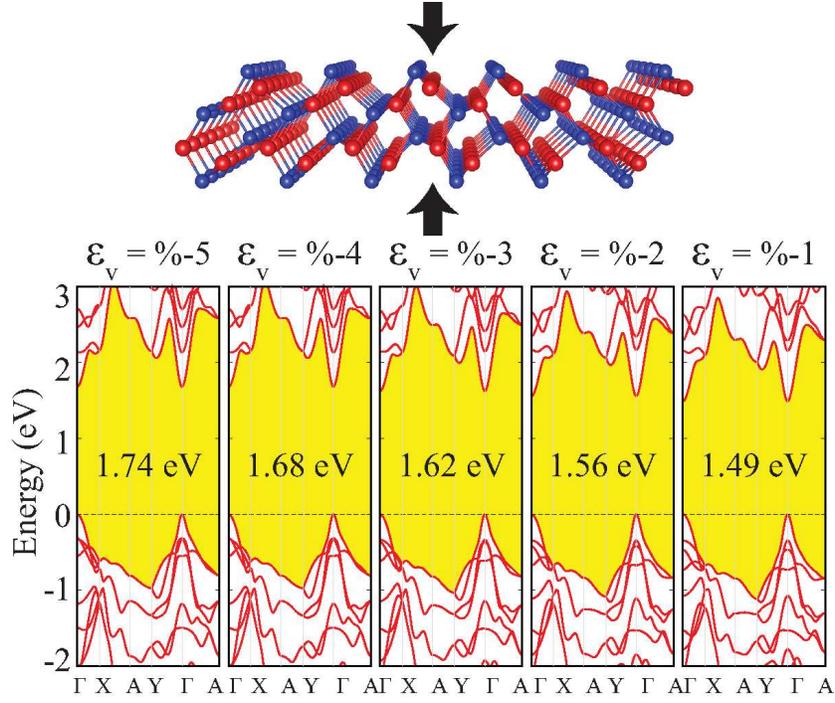}
\caption{\label{structure5}
(Color online) Evolution of the band dispersion of single-layer CdTe as a function of compressive strain 
along out-of-plane ($\varepsilon_{v}$) direction. Fermi level is set to zero.}
\end{figure}

\section{Strain Response of Single-Layer C\lowercase{d}T\lowercase{e}}
The built-in strain is inevitable as single-layer materials are usually grown on a substrate. It was shown that strain 
can significantly alter mechanical, electronic and magnetic properties of ultra-thin materials.\cite{Rodin, Jiang, Tao} 
Thus, in this section the effects of out-of-plane compressive, in-plane compressive and tensile strains on the 
direct-gap semiconducting behavior of single-layer CdTe are examined. The lattice constants of the unit cell for 
in-plane compressive and tensile strains are changed up to 5\% along zigzag (along $\overrightarrow{b}$) and armchair 
(along $\overrightarrow{a}$) directions. The thickness of the layer is compressed up to 5\% for out-of-plane compressive 
strain calculations.

\begin{figure}
\includegraphics[width=11 cm]{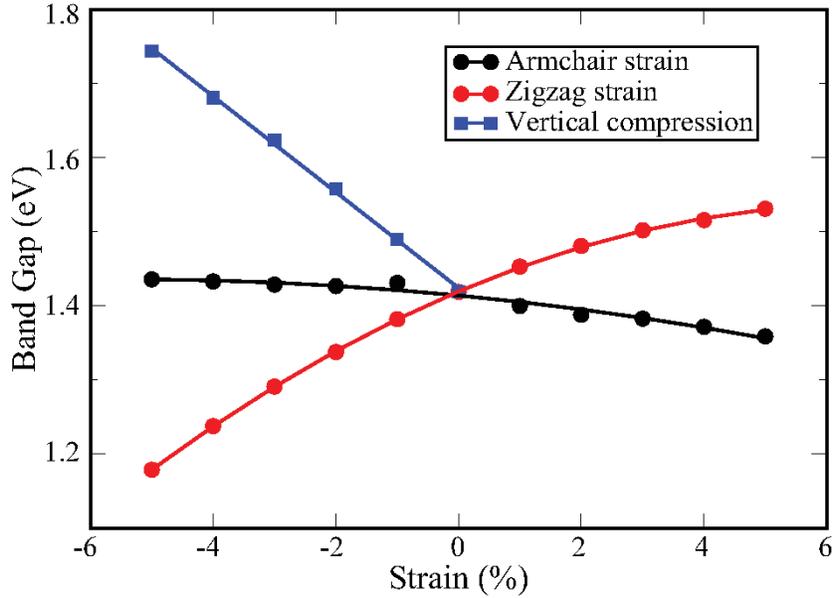}
\caption{\label{structure6}
(Color online) Evolution of band gaps of single-layer CdTe under in-plane and out-of-plane strain. Dots are 
calculated values and lines are fitted values.}
\end{figure}

Fig. \ref{structure4} illustrates the evolution of electronic band structures for strained CdTe single-layer. 
It is clearly seen that electronic characteristics of single-layer CdTe do not change significantly with 
applied in-plane strain. It exhibits robust direct-gap at the $\Gamma$ point under considered strain values. 
It is found that the band gap of single-layer CdTe is more sensitive to the in-plane strain applied along 
zigzag direction than armchair direction. With the increase of tensile strain along the armchair direction, 
the band gap of CdTe decreases, whereas the band gap increases when compressive strain along the armchair 
direction is increased. However, the increase of tensile strain along zigzag direction results in an increase 
in the band gap of CdTe, the increase of compressive strain leads to decrease in the band gap. Fig. 
\ref{structure5} shows electronic band structures for single-layer of CdTe under compression along 
out-of-plane direction. It is found that CdTe does not show significant structural distortion under 
considered out-of-plane compression values. Direct gap character of CdTe at the $\Gamma$ point does not 
change, but the electronic band gap increases as applied compressive strain increases.

Variation of the band gap of single-layer CdTe crystal under out-of-plane and in-plane strains are shown in 
Fig. \ref{structure6}. It was already calculated that VBM of the CdTe is mainly composed of \textit{p$_{y}$} 
orbitals of Te atoms. Since the Te-\textit{p$_{y}$} orbitals are aligned in the zigzag direction, 
modification of band edges via applied strain occurs much faster than those in armchair direction. As shown 
in Fig. \ref{structure6}, while the band gap slowly decreases with increasing strain in armchair direction, 
it rapidly increases with increasing strain in zigzag direction. 

Therefore, the variation of band gap of CdTe for applied tensile strain (in $\mp$5\%) along armchair 
direction is fitted to an expression as

\begin{equation}
 E_{gap} (\varepsilon_{arm})= 1.42 -\alpha \varepsilon_{arm}-\beta \varepsilon_{arm}^{2}
\end{equation}
$\alpha$ and $\beta$ are fitting parameters and their values are $\sim$ 0.008 and 0.001 eV. 
Compressive strain along the zigzag direction decreases the hybridization of \textit{p$_{y}$} 
orbitals of Te atom and \textit{d} orbitals of Cd atom at the VBM, whereas it increases the hybridization of 
Te and Cd orbitals at the CBM. Therefore, the VBM and CBM energies vary in opposite directions, thereby 
decreasing the band gap. The variation of band gap of single-layer CdTe for applied tensile strain along 
zigzag direction is fitted to an expression as

\begin{equation}
E_{gap} (\varepsilon_{zig})= 1.42 + \gamma \varepsilon_{zig}-\delta \varepsilon_{zig}^{2}
\end{equation}
where $\gamma$ and $\delta$ are fitting parameters and their values are $\sim$ 0.035 and 0.003 eV, 
respectively. As a result, strain-dependence of the band gap of single-layer CdTe exhibits nonlinear 
variations behavior when an in-plane strain is applied.

The out-of-plane strain application can easily alter the interlayer spacing of layered materials and 
therefore it provides an efficient way of tuning the electronic properties. In the Sec. III we found that the 
CBM of CdTe is dominated by \textit{p$_{z}$} orbitals of Te atom. Therefore, application of compressive 
out-of-plane strain significantly affects the hybridization between orbitals of Cd and Te. Consequently, the 
band gap of CdTe increases monotonically with increasing compressive strain along out-of-plane direction and 
the rate of change for band gaps is faster for the out-of-plane strain than that of in-plane strains. 
Increasing behavior of band gap of CdTe for applied compressive strain along out-of-plane direction is fitted 
to an expression as
\begin{equation}
E_{gap} (\varepsilon_{v})= 1.42 - \zeta \varepsilon_{v}
\end{equation}
$\zeta$ is a fitting parameter and it has a value of $\sim$ 0.065 eV. 

It appears that while the direct band gap feature is maintained, controllable modification of the band gap 
values of monolayer CdTe is feasible by the application of uniaxial strain along different crystallographic 
orientations. Mostly, electronic properties of ultra-thin materials are highly sensitive to the applied 
strain. It was shown that strain changes the energy dispersion, band gap, and the band edges of graphene.\cite{Wong} In 
another study, the optical band gap of MoS$_{2}$ experiences a direct-to-indirect transition with applied strain, which 
decreases the measured photoluminescence intensity.\cite{Conley} Previously we showed that electronic band structure of 
single-layer MoSe$_{2}$ undergoes a direct to indirect band gap crossover under tensile strain.\cite{Horzum} Moreover, 
strain induced phase transition (from semiconducting 2H phase to metallic 1T' phase) is observed in 
MoTe$_{2}$.\cite{Song} Therefore, in contrast to typical ultra-thin materials, monolayer CdTe exhibits robust and 
moderate band gap that covers the broad range of the solar spectrum, which are essential for its utilization in future 
electronics.

\section{Conclusions}

In this study, we investigated structural, phonon and electronic characteristics of single-layer CdTe by 
performing state-of-the-art first principle calculations. Structural analysis revealed that ultra-thin CdTe 
has a crystal structure made of reconstructed 8-atomic primitive unit cell. Electronic band dispersion 
calculations showed that single-layer CdTe has a direct band gap of 1.42 (GGA) eV at the $\Gamma$ point. Direction 
dependent energy band dispersions at the vicinity of VBM and CBM indicate that single-layer CdTe has 
anisotropic electronic and optical properties. 

Moreover, it is seen that electronic characteristics of single-layer CdTe are more sensitive to in-plane 
strain applied along zigzag direction than armchair direction. Along the armchair direction, the higher the 
tensile strain, the smaller the bandgap. However, increasing the tensile strain along zigzag direction 
increases the band gap. In addition, when a compressive strain applied in out-of-plane direction, the rate of 
increase of the electronic bandgap is much faster. It is also found that the direct bandgap semiconducting 
behavior of the ultra-thin CdTe is not affected by compressive and tensile strain applied in in-plane or 
out-of-plane directions. Ultra-thin CdTe crystal with its strain-independent and robust direct bandgap is 
quite suitable material for nanoscale optoelectronic device applications.

\end{document}